\documentclass{epl}

\usepackage{amsmath,amssymb,amsfonts,bm}
\usepackage{graphicx,eucal}

\title{Generalized entanglement as a natural framework for exploring quantum chaos}
\shorttitle{GE and quantum chaos}
\author{Y. S. Weinstein\inst{1}\thanks{Corresponding author. E-mail: \email{weinstein@mitre.org}}
 \and L. Viola\inst{2}\thanks{E-mail: \email{lorenza.viola@dartmouth.edu}}
}
\institute{
  \inst{1} Quantum Information Science Group, {\sc Mitre} - Eatontown, NJ 07724, USA\\
  \inst{2} Department of Physics and Astronomy, Dartmouth College - Hanover, NH 03755, USA
}

\begin{document}

\maketitle

\begin{abstract}
We demonstrate that generalized entanglement [Barnum {\em et al.},
Phys. Rev. A {\bf 68}, 032308 (2003)] provides a natural and reliable
indicator of quantum chaotic behavior. Since generalized entanglement
depends directly on a choice of preferred observables, exploring how
generalized entanglement increases under dynamical evolution is
possible without invoking an auxiliary coupled system or decomposing
the system into arbitrary subsystems.  We find that, in the chaotic
regime, the long-time saturation value of generalized entanglement
agrees with random matrix theory predictions.  For our system, we
provide physical intuition into generalized entanglement within a
single system by invoking the notion of extent of a state. The latter,
in turn, is related to other signatures of quantum chaos.
\end{abstract}

Central to the study of quantum chaos \cite{Haake} and broadly
significant to fundamental quantum theory \cite{PeresBook}, is the
determination of {\em distinctive} signatures that unambiguously
identify quantum systems whose classical limit exhibits chaotic,
versus regular, dynamics. Such signatures are discovered by
contrasting quantized versions of classically chaotic and non-chaotic
systems.  A well-established {\em static} signature of quantum chaos
is the accurate description of a chaotic operators' eigenvalue and
eigenvector element statistics by random matrix theory (RMT)
\cite{Haake,rmt}.  A {\em dynamic} indicator of quantum chaos is the
fidelity decay behavior \cite{Peres1,Jala,J1,P1,Jo,YSWRFD,review}.
While both approaches have led to deep insights into quantum chaos and
its relation to the underlying classical dynamics, they suffer from
intrinsic weaknesses.  Eigenvector statistics, for example, is
basis-dependent. The effectiveness of fidelity decay as an indicator
of quantum chaos is strongly influenced by the form of the
perturbation.  Indeed, regular systems may show chaotic fidelity decay
behavior depending on the type of perturbation \cite{Jo}.

A signature of quantum chaos which need not be subject to the above
weaknesses and is very natural from a quantum information standpoint
is entanglement generation.  Chaotic evolution tends to produce states
whose statistical properties are similar to those of {\em random} pure
states.  Because such states tend to be highly entangled \cite{rand},
we expect that quantum analogs of classically chaotic systems generate
greater amounts of entanglement than quantum analogs of non-chaotic
ones.  This has been confirmed both statically and dynamically.
Statically, by directly analyzing the entangling capabilities of the
evolution operator, and dynamically by studying the evolution of
specific initial states \cite{Lak,FNP,MS,TFM,J,Dem,WGSH,YSW}. However,
all of these studies require a preferred tensor product structure in
the ambient Hilbert space, in order for the standard definition of
entanglement to be applicable.  Thus, some of the above studies
arbitrarily decompose the system into subsystems \cite{WGSH,YSW},
while others couple the system to be studied to another system
\cite{Lak,FNP,MS,TFM,J,Dem}. The latter method introduces the coupling
strength as an extra degree of freedom, which can cause strongly
chaotic systems to {\em not} adhere to the proposed chaos indicator.
Ultimately, both of these methods effectively impose an external
architecture onto the system rather then studying the system on its
own terms.

A notion of {\em generalized entanglement} (GE) able to overcome the
limitations of the usual subsystem-based setting has been proposed in
\cite{L1}. GE extends the observation that standard entanglement can
be defined in terms of expectation values of a distinguished set of
observables, removing the need for a preferred subsystem
decomposition.  GE measures constructed from algebras of fermionic
operators have provided new diagnostic tools for probing many-body
correlations in quantum phase transitions \cite{L3}, and have
contributed to the understanding of standard multipartite entanglement
in disordered spin lattices \cite{Simone}.

In this Letter, we establish GE production with respect to appropriate
observable sets as an indicator of quantum chaos which removes the
above-mentioned weaknesses.  In particular, because the GE framework
relies only on {\it convex} structure of the spaces of quantum states
and observables, GE is able to be defined within the system {\em
alone}, without resorting to coupling additional systems or imposing
arbitrary subsystems.  We demonstrate how GE clearly differentiates
between fully chaotic, partially chaotic, and regular behavior using
the paradigmatic case of a {\em quantum kicked top} (QKT) \cite{H2}.
Furthermore, we show that the behavior of the chaotic QKT follows the
RMT prediction. Finally, we provide a physical justification by
comparing GE to the notion of {\em extent} of a state, introduced by
Peres \cite{Peres2}, and recently related to fidelity decay
\cite{YSWRFD}.
 

The starting point to define GE is to realize that standard entangled
pure states of a composite quantum system $S$ look mixed to observers
whose means to control and measure $S$ are constrained to local
operations on individual subsystems: To specify a pure entangled state
requires knowledge of the correlations, which are expectations of
non-local operators.  By thinking of pure states as one-dimensional
(extremal) projectors in the set of density operators for $S$,
entanglement implies a loss of purity (extremality) upon restricting
to local expectations only.  A similar characterization can be
provided without making reference to a subsystem decomposition for
$S$.  Let $S$ be defined on a Hilbert space ${\cal H}$, and let
$\Omega$ denote a {\em generic} set of observables.  Then any pure
state $|\psi\rangle \in {\cal H}$ induces a {\em reduced} state that
determines only the expectations of operators in $\Omega$.  In analogy
with the standard case, $|\psi\rangle$ is said to be {\em generalized
unentangled relative to $\Omega$} if its reduced state is pure,
generalized entangled otherwise \cite{L1}.

A natural way to quantify GE is to relate $|\psi\rangle$ to $\Omega$
via the (square) length of the projection $|\psi\rangle\langle \psi|$
onto $\Omega$ \cite{L1}. We shall focus on the case where $\Omega
\equiv \mathfrak{h}$ is a real Lie algebra faithfully represented on
${\cal H}$, linearly spanned by a set $\{A_\ell\}$, $\ell=1,\ldots,
L$, of Hermitian operators, orthogonal with respect to the trace norm
\cite{L3}. The {\em purity of $|\psi\rangle$ relative to}
$\mathfrak{h}$ ($\mathfrak{h}$-purity) is
\begin{equation}
P_{\mathfrak{h}}(|\psi\rangle) = {\tt K} \sum^L_{\ell= 1}\langle \psi
| A_{\ell}|\psi \rangle^2 = {\tt K} \sum^L_{\ell= 1}\langle
A_{\ell} \rangle^2\:,
\label{purity}
\end{equation}
where the constant ${\tt K}>0$ ensures that the maximum value of
$P_{\mathfrak{h}}$ is $1$. States of maximum purity are generalized
unentangled relative to $\mathfrak{h}$. If the latter is a semisimple
Lie algebra acting irreducibly on ${\cal H}$, then any generalized
unentangled state has extremal length, and belongs to the family of
{\em generalized coherent states} (GCSs) \cite{Perelomov}.


The quantum chaotic system we explore is the QKT, used in many
previous studies of quantum chaos in general \cite{Haake} and
entanglement generation in particular. In contrast to the present
work, however, previous studies of QKT entanglement either explored
coupled kicked tops \cite{Lak,Dem}, or a realization of the QKT in
term of spin $1/2$ susystems \cite{WGSH}. The dynamical variables of
the QKT are the three components of the angular momentum vector, ${\bf
J}=(J_x,J_y,J_z)$, with $|{\bf J}|=J$ constant.  The dynamics of the
classical kicked top is a locus of points on the surface of the unit
sphere spanned by ${\bf J}/J$, with the relative size of the
non-chaotic and chaotic regions depending on the kick strength,
$k$. The kicked top is fully non-chaotic for $k \lesssim 2.7$, has
both chaotic and non-chaotic regions for $2.7 \lesssim k \lesssim
4.2$, and is fully chaotic for $k \gtrsim 4.2$ \cite{J1}.  QKT
evolution is generated by the Floquet operator \cite{H2}
\begin{equation}
U_{\tt QKT} = e^{-i\pi J_y/2}e^{-ik J_z^2/2J},\;\;\hbar =1\,,
\end{equation}
in a Hilbert space ${\cal H}_N$ of dimension $N = 2J+1$. ${\cal H}_N$
furnishes a spin-$J$ irreducible representation of SU(2), thus, it is
natural to investigate $\mathfrak{h}=\mathfrak{su}(2)$ as a preferred
algebra for this system.  From Eq.~(\ref{purity}), the
$\mathfrak{su}(2)$-purity is
\begin{equation}
P_{\mathfrak{su}(2)}(|\psi\rangle) = \frac{1}{J^2} \sum_{\ell=x,y,z}
\langle \psi |J_\ell| \psi \rangle^2 \:,
\label{P}
\end{equation}
where ${\tt K}=J^{-2}$ is chosen so that $P_{\mathfrak{su}(2)} = 1$
for angular momentum GCSs, defined by the eigenvalue equation $({\bf
n}\cdot {\bf J}) |\psi\rangle = J |\psi\rangle$, ${\bf n}=(\sin\theta
\cos\phi, \sin\theta \sin\phi, \cos\theta)$, $\theta \in [0,\pi],
\phi\in [-\pi, \pi]$ \cite{PeresBook}.  Thus, GE$_{\mathfrak{su}(2)} =
1 - P_{\mathfrak{su}(2)}$. 


In the chaotic regime, RMT predicts the asymptotic state of the QKT to
be described by a random pure state uniformly drawn according to the
Haar measure on SU($N$).  Following the general procedure for
estimating the expected GE in typical pure states \cite{winton}, or
exploiting the fact that the above expectation values have been
previously studied within RMT \cite{Haake,Dem}, the average
$\mathfrak{su}(2)$-GE is found to be
$\overline{\mbox{GE}}_{\mathfrak{su}(2)} = 1 -1/2J.$

\begin{figure}
\onefigure[width=8cm]{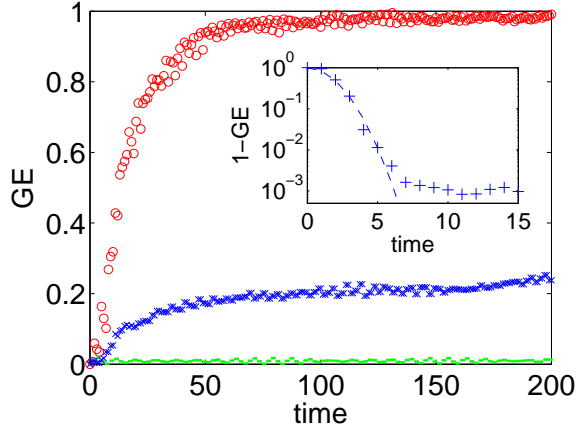}
\caption{GE$_{\mathfrak{su}(2)}$ versus time for representative
initial angular momentum GCSs under the evolution of a mixed phase
space QKT, $k = 3$, $J = 500$. The GE$_{\mathfrak{su}(2)}$ of the GCS
centered in a chaotic region $\theta = 3\pi/5, \phi = -\pi/10$
($\bigcirc$), quickly approaches one. The state centered in the
regular region, $\theta = \pi/2, \phi = 0$, generates very little
GE$_{\mathfrak{su}(2)}$ ($\cdot$), and the state centered at the edge
of chaos $\theta = \pi/2, \phi = -\pi/10$, exhibits intermediate
behavior ($\times$).  Inset: Average $\mathfrak{su}(2)$-purity of $90$
GCSs under chaotic QKT evolution, $k = 12$, $J=500$. The
GE$_{\mathfrak{su}(2)}$ increases as a Gaussian, $e^{-0.18t^2}$
(dashed line), saturating at $\approx 0.999$, the RMT estimation.  }
\label{QKTM1}
\end{figure}

We begin by exploring a QKT with a mixed phase space, $k =
3$. Fig.~\ref{QKTM1} contrasts the GE$_{\mathfrak{su}(2)}$ growth as a
function of time for GCSs centered in the chaotic versus regular
region of the classical phase space. States in the chaotic region
quickly approach the GE$_{\mathfrak{su}(2)}$ value predicted by RMT,
whereas states in the regular region generate much less GE.  GCSs at
the ``edge of quantum chaos'' \cite{YSW2}, the border between the
chaotic and regular phase space regions, demonstrate intermediate
behavior.

\begin{figure}
\twoimages[width=6.5cm]{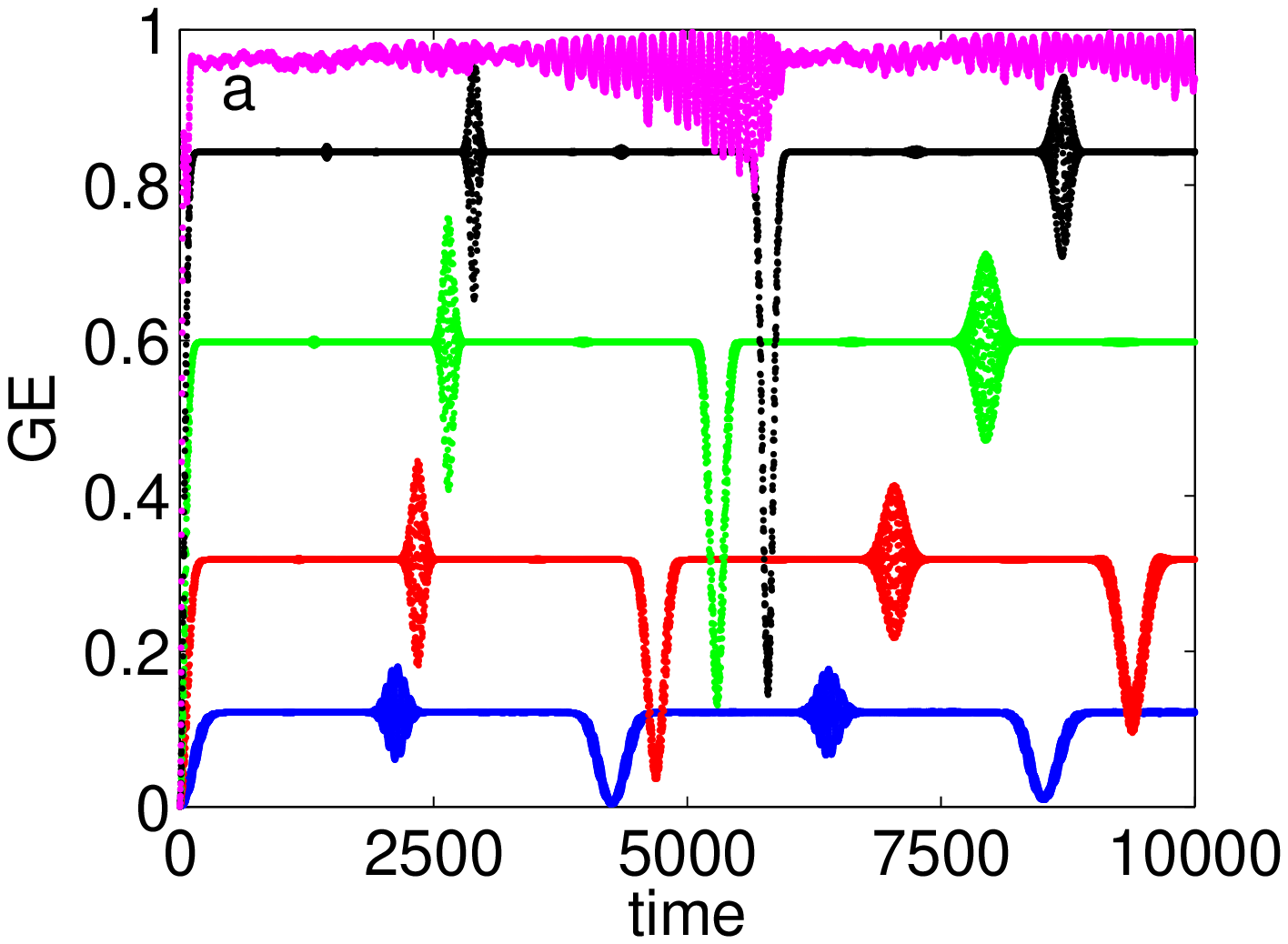}{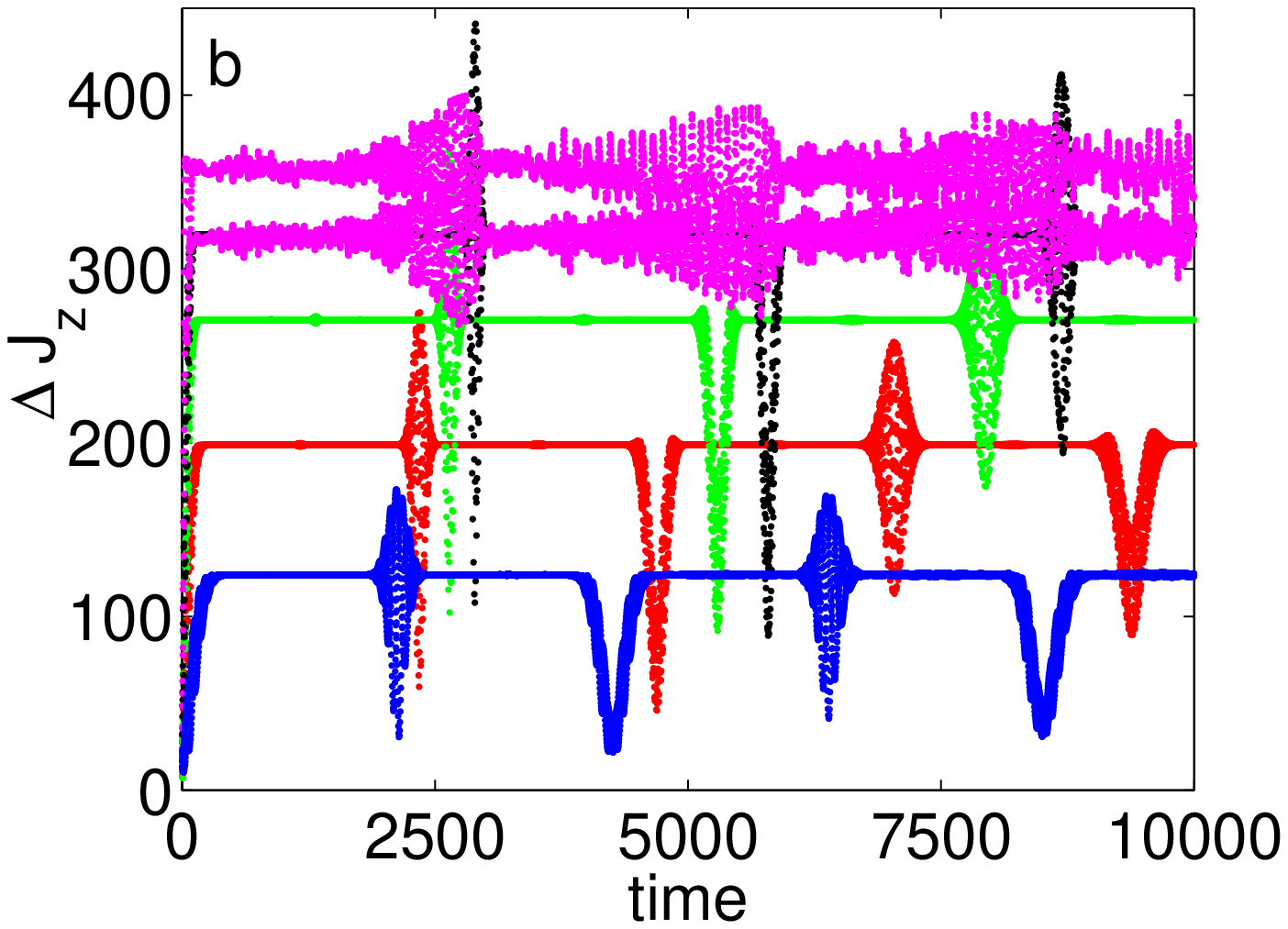} 
\caption{GE$_{\mathfrak{su}(2)}$ (a), and $z$-extent, $\Delta J_z$ (b)
for representative initial angular momentum GCSs under regular QKT
evolution, $k = 1.1$, $J = 500$.  The states (bottom to top) are
centered at $\theta = 3\pi/5, \phi = -2\pi/5, -3\pi/10, -\pi/5,
-\pi/10, 0$, which lie on phase space orbits of increasing size.  The
first four states exhibit linear GE and extent increase, until
saturation at a level which depends on the size of the GCS orbit.
When the orbit is large, the GCS has a larger spread with respect to
$\mathfrak{su}(2)$-observables, hence a larger GE and $\Delta
J_z$. After saturation, periodic recurrences in both the GE and extent
are seen.  The {\em initial} extent is a good indicator of regular
regime fidelity decay behavior \cite{YSWRFD}.  The above four GCSs are
mainly composed of a few low-extent eigenvalues and show a Gaussian
fidelity decay.  The GCS $\theta = 3\pi/5, \phi = 0$, does not display
any recurrences. Rather, both the GE and extent exhibit wild
oscillations and achieve higher values than the other states.  This
state is composed of several high-extent QKT eigenvalues and exhibits
a power-law fidelity decay under small perturbations
\protect\cite{YSWRFD}.  }
\label{SG}
\end{figure}

As $k$ increases, the chaotic sea covers the whole of phase space.
Correspondingly, the GE$_{\mathfrak{su}(2)}$ of {\em all} states
quickly approach the RMT estimation.  The inset of Fig.~\ref{QKTM1}
illustrates this for a QKT of $k = 12$.  The GE$_{\mathfrak{su}(2)}$
initially increases as a Gaussian and then plateaus at $0.999$, as
predicted. In contrast, a QKT with a regular phase space, $k = 1.1$,
displays an initial linear GE$_{\mathfrak{su}(2)}$ growth that
typically plateaus well below one, Fig.~\ref{SG}(a).

The above results strongly support the use of GE as a signature of
quantum chaos but do not offer a clear insight into the physical
meaning of this property in our system.  To clarify this concept we
establish a relationship between GE and the extent of a state relative
to a Hermitian observable $A$ \cite{Peres2,note}. The latter is
defined as
\begin{equation}
\Delta A(|\psi\rangle) = \sqrt{\langle A^2\rangle-\langle A\rangle^2}\:.
\label{ext}
\end{equation}
Thus, the extent is the square-root of the variance of $A$ for the
state $|\psi\rangle$.  The connection to GE is shown by noting that,
for $\mathfrak{h}$ irreducible, the $P_{\mathfrak{h}}$-purity is
directly related to the {\em invariant uncertainty} functional,
$(\Delta {\cal I})^2$. With respect to the above operator basis \cite{L3},
\begin{equation}
(\Delta {\cal I})^2 (|\psi\rangle)= \sum_{\ell}[\langle
A_{\ell}^2\rangle - \langle A_{\ell}\rangle^2] = \langle C_2\rangle -
P_{\mathfrak{h}}(|\psi\rangle)\:,\label{inv}
\end{equation}
where $C_2$ is the quadratic Casimir invariant of ${\mathfrak{h}}$,
here $C_2={\bf J}\cdot {\bf J}$.  Thus, Eqs.~(\ref{ext})-(\ref{inv})
yield
\begin{equation}
\mbox{GE}_{\mathfrak{su}(2)}= \sum_{\ell=x,y,z} 
\Delta \Big( \frac{J_\ell}{J}\Big)^2 -\frac{1}{J}\:.
\label{gext}
\end{equation}
Eq.~(\ref{gext}) clarifies how GE relative to the angular momentum
observables is directly related to (identical to, as
$J\rightarrow\infty$) the squared extent for rescaled observables
$J_\ell/J$, {\em averaged} over $x,y,z$.  As suggested in
\cite{PeresBook,Peres2}, the extent in each direction contains
essentially equivalent information for differentiating between regular
and chaotic dynamics. For the QKT, this may be seen explicitly: due to
the $\pi/2$ rotation in $U_{\tt QKT}$, $x$ and $z$ are interchanged at
every time step, leading to equivalent $z$- and $x$-extent
behavior. The $y$-extent is bounded by $(\Delta J_y)^2 \leq J(J+1)-2
(\Delta J_z)^2$ causing large $x,z$-extents to be correlated with
small $y$-extent values.  Thus, the behavior of
GE$_{\mathfrak{su}(2)}$ as a quantum chaos indicator should be {\em
qualitatively similar} to the extent behavior for any observable
$J_\ell/J$.

The relation between the $z$-extent and GE$_{\mathfrak{su}(2)}$ is
exhibited graphically in Fig.~\ref{SG} using initial GCSs under
regular QKT evolution, $k = 1.1$.  The similarity is striking: Both
increase linearly as a function of time until some saturation
level. Upon saturation, both the extent and GE$_{\mathfrak{su}(2)}$
plateau, except for periodic recurrences which occur at the same time.
An analogous behavior has been numerically verified
for extents in the $x,y$ directions.

The relationship between GE$_{\mathfrak{su}(2)}$ and the extent
provides the following intuitive, physical picture for GE and
``self-entanglement'': GE$_{\mathfrak{su}(2)}$ is analogous to a
measure of the spread of the system's state vector in the phase space
associated with the dynamical observables.  As an initial GCS evolves
to cover more and more of its orbit, the GE$_{\mathfrak{su}(2)}$
grows.  The larger the phase space orbit of the GCS, the larger the
entanglement saturation level (Fig. \ref{SG}). For fully chaotic
systems, any typical orbit covers all of phase space, hence the
GE$_{\mathfrak{su}(2)}$ converges to the RMT estimation. In this
sense, GE$_{\mathfrak{su}(2)}$ evolution, {\em at least starting from
states which have a good classical limit}, directly reflects the
underlying classical phase space structure. Similar connections
between the entanglement growth and the spread of an initial GCS have
been made for standard bi-partite entanglement, the rate of
entanglement increase being determined by the Lyapunov exponents of
the corresponding classical Liouville distribution
\cite{MS,J,Dem,PJ}. Additional insight into single-particle entanglement
has been provided in \cite{Can}.

As mentioned, previous studies of QKT entanglement relied on
decomposing the system into $N = 2J$ spin $1/2$ systems and
restricting to states symmetric under spin exchange \cite{WGSH}. It
can then be shown that GE$_{\mathfrak{su}(2)}$ is {\it equivalent} to
standard global multipartite entanglement as quantified by the
Meyer-Wallach measure \cite{MW} that is, GE$_{\mathfrak{su}(2)}$ is
proportional to the average linear entropy of entanglement between any
spin $1/2$ subsystem and the rest.  Yet, the classical picture of
spread in phase space is still useful in showing the close
relationship between chaos and entanglement generation. Formally, the
connection between GE$_{\mathfrak{su}(2)}$ and other entanglement
measures demonstrates how GE unifies different entanglement
approaches.


Through the extent, GE$_{\mathfrak{su}(2)}$ is connected to fidelity
decay -- a quantum chaos signature which provides a fingerprint of the
classical Lyapunov exponent for quantized versions of classically
chaotic systems \cite{Jala,C1}. Fidelity is a measure of distance
between the states reached from a given initial state $|\psi_i\rangle$
under slightly different evolutions \cite{Peres1}, $F(t) =
|\langle\psi_i|U^{-t}U_p^t|\psi_i\rangle|^2$, where $U^t=e^{-iH_0t}$,
$U_p^t = e^{-i(H_0+\delta V)t}$ are the unperturbed and perturbed
evolutions, and $\delta$, $V$ are the perturbation strength and
Hamiltonian, respectively.  At short time, $F(t) = 1 - (\langle V^2
\rangle - \langle V \rangle^2)\delta^2t^2 + ... \,,$ immediately
identifying the square of the {\em initial} $V$-extent as the
coefficient of the second-order term \cite{Peres2}.  While a complete
characterization of the operators $V$ able to induce effective
dynamical cross-over is lacking \cite{Jo}, perturbations commuting
with one of the $J_\ell$ make fidelity a reliable indicator for QKT
dynamics \cite{PeresBook,J1}. This relation between the $\ell$-extent
and fidelity decay further supports the validity of
GE$_{\mathfrak{su}(2)}$ as a quantum chaos indicator (see also
Fig.~\ref{SG}). We note that fidelity decay of a GCS is also related
to the size of the classical phase space orbit \cite{YSWRFD}. In
general, fidelity decay is connected to other signatures of quantum
chaos, such as the shape of the local density of states \cite{J1} and
eigenvector statistics \cite{Jo}. Montangero {\it et al} \cite{MBF}
demonstrate the similarity of behavior between fidelity decay and the
decay of (bi-partite) entanglement of an initial Bell pair.  Here, it
is the generalized purity which decays and, as shown, behaves
qualitatively similarly to fidelity decay -- complementing the results
obtained for local purity and fidelity in \cite{Simone}.

The fact that a single spatial direction suffices for identifying a
valid fidelity perturbation and extent variable suggests that we
examine an observable set consisting of a {\em single observable} as
another candidate for defining GE.  This is done by restricting to a
(Cartan) subalgebra $\mathfrak{h}=\mathfrak{so}(2) \subset
\mathfrak{su}(2)$ generated by a single operator $J_\ell$, say
$J_z$. Using Eq.~(\ref{purity}), the $\mathfrak{so}(2)$-purity is
\begin{equation}
P_{\mathfrak{so}(2)}(|\psi\rangle) = \frac{1}{J^2}
\langle \psi |J_z| \psi \rangle^2 \,, \;\;
\mbox{GE}_{\mathfrak{so}(2)}=1 - P_{\mathfrak{so}(2)}\,.
\end{equation}
Numerical simulations for initially ${\mathfrak{so}(2)}$-unentangled
GCSs ($\theta=0$) show that GE$_{\mathfrak{so}(2)}$ is also a valid
signature of quantum chaos for this system (data not shown).  This
indicates the ability of the $\mathfrak{h}$-purity to differentiate
between regular and chaotic dynamics without a direct link to
variances of observables or invariant uncertainty.  Suggestively, the
${\mathfrak{so}(2)}$-purity has been shown to characterize quantum
criticality in the Lipkin-Meshov-Glick model~\cite{L3}, which may also
be mapped into a single (pseudo)spin system with ${\mathfrak{su}(2)}$
dynamical algebra.

The above discussion shows the utility of expectation values and
statistical moments of observables in understanding quantum chaos.
Uncertainty-based entanglement measures have been suggested outside 
the GE framework \cite{DDJ} providing
links to important quantities such as the Wigner-Yanasi skew
information \cite{ZChen,Can} and a quantum analog of the Fisher
information \cite{Luo}.  A dedicated study connectng GE to such 
measures will be presented elsewhere. A {\em general}
characterization of a preferred set of observables which can sharply
differentiate between chaotic and non-chaotic regimes likewise remains
an area for future in-depth analysis.  In particular, this will
require extending the present study to other quantum chaos models --
for instance, kicked rotors and the quantum baker's maps.  While 
further generalizations of the mathematical
formalism are likely to be needed (in order to properly define, for
example, GCSs for {\em discrete} groups), we believe that the GE
notion has both the flexibility and the potential for identifying 
quantum chaos signatures in arbitrary physical settings.

We thank S. Ghose, G. Gilbert, C.S. Hellberg, S. Montangero and
L.F. Santos for feedback.  Y.S.W. acknowledges support from the
National Research Council through the NRL and the MITRE Technology 
Program Grant No. 07MSR205. Some
computations were performed at the ASC DoD Major Shared Resource
Center. L.V.  acknowledges partial support from C. and W. Burke
through their Special Projects Fund in Quantum Information Science.


\begin{thebibliography}{99}
\bibitem{note} In the original definition \cite{Peres2}, Peres is
interested in states within a fixed symmetry class thus the variance
of $|A|$ is considered.
Since (as in \cite{YSWRFD}) we are not restricted in this way, we
remove the absolute value in our analysis.
\end{thebibliography}
\end{document}